\documentclass[onecolumn]{IEEEtran}
\usepackage{cuted}%%\stripsep-3pt
\usepackage{cite}
\usepackage{amsthm}
\usepackage{amsmath}
\usepackage{comment}
\usepackage{booktabs}
\usepackage{bm}
\usepackage{setspace}
\usepackage{graphicx}

\usepackage[usenames,dvipsnames]{color}
\usepackage{amsfonts,amssymb,amsmath}
\usepackage{subfigure}
\usepackage{enumerate}
\usepackage{cases}
\usepackage{array}
\usepackage{colortbl}
\usepackage{multirow}
\usepackage{arydshln}
\usepackage{soul, color, xcolor}
\usepackage{float}

\soulregister{\cite}7 % 注锟斤拷\cite锟斤拷锟斤拷
\soulregister{\citep}7 % 注锟斤拷\citep锟斤拷锟斤拷
\soulregister{\citet}7 % 注锟斤拷\citet锟斤拷锟斤拷
\soulregister{\ref}7 % 注锟斤拷\ref锟斤拷锟斤拷
\soulregister{\pageref}7 % 注锟斤拷\pageref锟斤拷锟斤拷
\newtheorem{thm}{Theorem}

\newtheorem{lem}{Lemma}

\theoremstyle{remark}

\definecolor{Gray}{gray}{0.85}
\definecolor{mycyan}{cmyk}{.3,0,0,0}

\newcolumntype{a}{>{\columncolor{Gray}}c}%\centering\backslash
\newcolumntype{b}{>{\columncolor{white}}c}
\newcolumntype{d}{>{\columncolor{mycyan}}c}

\allowdisplaybreaks
\begin{document}
%{ Violation Probabilities of AoI and PAoI and Optimal Arrival Rate Allocation for the IoT-based Multi-Source Status Update System}
\title{
%\LARGE
{Age of Information in a Multisource Ber/Geo/1/1 Preemptive Queueing System}}
%\title{On the Timeliness of the Stalest Information: Age of the Stalest Information and Peak Age of the Stalest Information}
% Authors, for the paper (add full first names)
\author{{Tianci~Zhang and Zhengchuan~Chen}
% Zhong~Tian,~\IEEEmembership{Member,~IEEE,} Yunjian~Jia,~\IEEEmembership{Member,~IEEE,} Min~Wang,~\IEEEmembership{Member,~IEEE,} and Tony~Q.~S.~Quek,~\IEEEmembership{Fellow,~IEEE}
%\vspace{-9 mm}
%%and Dapeng~Oliver~Wu,~\IEEEmembership{Fellow,~IEEE}
\thanks{Tianci Zhang and Zhengchuan Chen are with the School of Microelectronics and Communication Engineering, Chongqing University, Chongqing 400044, China (e-mails: \{ztc, czc\}@cqu.edu.cn).}}
\maketitle

\begin{abstract}
This work studies the information freshness of the
%vehicle-to-vehicle and
vehicle-to-infrastructure status updating in Internet of vehicles, which is modeled as a multi-source Ber/Geo/1/1 preemptive queueing system with heterogeneous service time.
%To strictly characterize and validly evaluate the freshness, w
We pay attention to both the
%probability mass function (p.m.f.)
distribution
and average of AoI. To fully track the per-source AoI evolution, a Markov two-dimensional (2D) age process is introduced.
The first element of the 2D age process stands for the instantaneous per-source AoI, while the second represents whether an update of the concerned source is being served and its current age.
A complete
% analysis
framework
and detailed analyses
on the per-source AoI are presented based on the Markovity of the 2D age process.
%An explicit framework for analysing the per-source AoI process by utilizing the 2D age process is presented.
By studying the state transition probabilities, stationary equations, and stationary distribution of the 2D age process, analytical expressions of the probability mass function and average of per-source AoI are derived.
Numerical results validate the accuracy of the theoretical analyses.
\end{abstract}
%\vspace{-2.5 mm}
%\vspace{-2.5 mm}
\section{Introduction}\label{intro}
Internet of vehicles (IoV)
%, as a significant application of Internet of things (IoT),
is in accelerating development with the rapid advances of sensing, communication, computation and artificial intelligence technologies.
In IoV, the
%vehicle-to-vehicle (V2V) and
vehicle-to-infrastructure (V2I) communication
provides the real-time vehicle status updating,
facilitating the  timely
%vehicle-status-information-based
vehicular-data-driven
 analysis and decision-making at the roadside infrastructure (RI)\cite{iovv2vv2i}.
In effect, the
% V2V and
V2I link is usually extracted as a real-time wireless status updating system consisting of multiple information sources, a transmitter and a monitor\cite{9316802}.
Information freshness becomes one of the critical indicators in the V2I link.
Age of information (AoI), which is defined as the time elapsed since the latest successfully received update of a concerned source was generated, is pertinent to characterize the information freshness for the status updating system\cite{9380899}.
%\cite{9380899}.

Queueing systems are effective to model the wireless status updating systems, providing abundant insights for improving the information freshness\cite{cstiov,8006504,8406928,9718306,tripathi2019age,9312180,8820073,tmd,woo,9965796,9611393,9377627,9333607}.
%The AoIs of the single-source queueing systems were first studied in seminal works\cite{}.
%In particular, c
In particular, multi-source queueing systems are more appropriate and accurate for modeling the status updating of the V2I link, since there are often multiple kinds of vehicular status, e.g. velocity, acceleration and surrounding object information, required to report to the RI\cite{cstiov}.
%%compared to single-source queueing systems,
%multi-source queueing systems can model the
%%the V2V or
%V2I link
%with more validity,
%%more validly,
%since the vehicle often updates many kinds of vehicle status information, e.g., velocity,  acceleration and surrounding object to
%%another vehicle or
%the RI
%% via a transmitter
%\cite{cstiov}.
%%To improve the information freshness, some packet management policies, e.g., preemption and non-preemption queuing disciplines were introduced\cite{8006504,8406928,9718306}.
%, which motivates the AoI analysis in the preemptive and non-preemptive queueing systems\cite{8006504,8406928,9718306}.
%Based on the queuing theory,
For queueing systems, most existing works adopted the average AoI, i.e, expectation or time-average of the AoI process to evaluate the information freshness\cite{8006504,8406928,9718306,tripathi2019age}.
%Average AoI, i.e., the expectation or time-average of the AoI process is commonly adopted for evaluating the information freshness\cite{8006504,8406928,9718306,tripathi2019age}.
%Information freshness can be appropriately evaluated by the average AoI, which is the time-average or expectation of AoI process\cite{8006504,8406928,9718306,tripathi2019age}.
%The average AoIs in the single-source M/D/1, D/M/1 and M/M/1 first-come-first-served (FCFS) systems were first found\cite{6195689}.
%The authors of \cite{8006504} derived the average AoIs in the single-source M/G/1/1 preemptive and non-preemptive system.
%Three approximate expressions of that in the multi-source M/G/1 FCFS system were found in \cite{9217386}.
For instance, average AoIs in the single- and multi-source M/G/1/1 preemptive and non-preemptive systems were derived in \cite{8006504,8406928,9718306}.
For discrete-time queueing systems,
%the explicit expressions of
the average AoIs in single-source Ber/G/1 first-come-first-served (FCFS), G/G/1/1 preemptive and G/G/$\infty$ systems were established in \cite{tripathi2019age}.
%Moreover, to improve the information freshness, some packet management policies, e.g., preemption and non-preemption queuing disciplines were introduced\cite{8006504,8406928,9718306}.

To characterize the information freshness comprehensively, some recent works focused on studying the (stationary) distribution of AoI\cite{9312180,8820073,tmd,woo,9965796,9611393,9377627,9333607}.
For continuous-time queueing systems, the methods for deriving the distribution of AoI mainly include the stochastic hybrid systems (SHS) approach \cite{9312180} and the sample-path (SP) approach\cite{8820073,tmd,woo}.
%The authors of \cite{8820073} analysed the distributions of AoI in the continuous-time single-source G/G/1 FCFS as well as G/G/1/1 preemptive and non-preemptive systems by using the sample-path approach.
%With the help of the sample-path approach,
%In particular,
The distributions of AoIs in the single- and multi-source M/G/1/1 preemptive and non-preemptive systems were investigated by using the SP approach in\cite{8820073,tmd,woo}.
In contrast, the two-dimensional (2D) age process approach\cite{9965796,9611393} and the SP approach\cite{9377627} are two main methods for analysing the distributions of AoI in discrete-time queueing systems.
%for discrete-time
%%queueing
%systems.
In particular, the 2D age process introduces a 2D Markov age process to fully track the AoI evolution, which can be seen as a generalization of the SHS approach\cite{9611393}. The first element of the 2D age process stands for the age process at the monitor, while the second represents the age of the update being served.
%This approach can be seen as a generalization of the SHS approach.
%The Markovity of the AoI process in discrete-time system is sufficiently utilized in this approach.
Compared with the SP approach, the AoI evolution is presented more visually by utilizing the 2D age process approach, facilitating to study the statistical characteristics of the AoI thoroughly.
%where
%The distributions of AoI in the single-source Ber/Geo/1 FCFS and Ber/Geo/1/1 preemptive systems were obtained by using the SP approach in\cite{9377627}.
With the help of the 2D age process, the distributions of AoIs in the single-source Ber/Geo/1 FCFS and Ber/G/1/1 non-preemptive systems were derived in \cite{9965796,9611393} respectively.
%The authors of \cite{9965796,9611393}
%With the introduction of the 2D age process approach where a Markov 2D age process is utilized to fully characterize the AoI process, the distributions of AoI in the single-source Ber/Geo/1 FCFS and Ber/G/1/1 non-preemptive systems were derived in \cite{9965796,9611393}.
%The first element of the 2D age process represents the AoI process, while the second element is defined as the age of the update being served.
%This approach can be seen as a generalization of the SHS approach.
Moreover, an algorithm based on the discrete-time Markov chain (DTMC) theory
% of quasi-birth-death type
was proposed to numerically obtain the distributions of per-source AoIs in multi-source bufferless preemptive and non-preemptive systems with  Bernoulli arrivals and discrete phase-type service times\cite{9333607}.
%In effect, the AoI in the discrete-time multi-source queueing system was rarely studied

%using the distribution and average of AoI
%It is necessary to
Thorough studies on the information freshness in the V2I link are necessary due to highly dynamic nature of the vehicle status.
%, and the distribution and average AoI are  the foundations.
%It is necessary to thoroughly st
%strictly characterize and evaluate the information freshness in the V2I link with the help of the distribution and average of AoI, due to the high dynamism of the vehicle status.
%Information freshness in the V2I link is worthy
%
%As aforementioned, information freshness is critical for the
%%V2V or
%V2I link due to the high dynamism of the vehicle status, and the distribution and average of AoI strictly characterize and evaluate the freshness.
%As far as we know,
However,to the best of our knowledge, the statistical characteristics of AoI (which are key to many extended studies\cite{woo})
%i)
in the discrete-time multi-source queueing system (which can model the
%V2V or
V2I link effectively) have not been sufficiently studied.
%; ii) The analytical results of the distribution and average of AoI in the multi-source Ber/Geo/1/1 preemptive system  has not been found.
%Motivated by these, we in this work strictly analyse the per-source AoI process in a multi-source Ber/Geo/1/1 preemptive system with the source-dedicated service time by using the 2D age process approach. The Markovity of the 2D age process helps overcome the challenge to analyse the frequently disturbed queuing process caused by the preemption among sources.
%%In particular, analytical expressions of the distribution and average of per-source AoI are derived.
%%we study the distribution and average of AoI in the multi-source Ber/Geo/1/1 preemptive system with the source-dedicated service time.
%The main contribution are summarized as follows.
%\begin{itemize}
%  \item We extend the 2D age process approach into the multi-source system.
%  To fully track the per-source AoI evolution, we define the first element of the 2D age process as the instantaneous per-source AoI, as well as utilize the second element to represent whether an update of the concerned source is being transmitted and its current age.
%In particular, we adopt the age evolution model where the minimum of AoI is two slots, which help describe the
%  \item
%\end{itemize}
To fill this gap, we study the AoI in a V2I link with multiple information sources, a vehicle of bufferless transmitter, retransmission protocol and preemption queueing policy, which is modeled as a multi-source Ber/Geo/1/1 preemptive system with the source-dedicated service time, and focus on the probability mass function (p.m.f.) and average of AoI.
%Specifically, to fully track the per-source AoI evolution, we utilize the 2D age process approach and extend it into the multi-source system.
%%we extend the 2D age process approach to analyse the per-source AoI process, where
%The Markovity of the 2D age process helps overcome the challenge to analyse the frequently disturbed queuing process caused by the preemption among sources.
%The main contributions of this work can be summarized as follows.
%\begin{itemize}
%  \item It is found that for the multi-source Ber/Geo/1/1 preemptive system, slightly
%  \item
%  \item
%\end{itemize}
%To fill this gap, we
%in this work
%\begin{itemize}
%  \item extend the 2D age process approach into the multi-source system and present a complete framework;
%  \item strictly analyse the per-source AoI process in a multi-source Ber/Geo/1/1 preemptive system with the source-dedicated service time;
%  \item first derive the analytical expressions of the probability mass function (p.m.f.) and average of the per-source AoI;
%  \item and analyse the average AoI versus all system parameters.
%\end{itemize}
%To fill this gap, we
%i) extend the 2D age process approach into the multi-source system and present a complete framework; ii) strictly analyse the per-source AoI process in a multi-source Ber/Geo/1/1 preemptive system with the source-dedicated service time; iii) first derive the analytical expressions of the probability mass function (p.m.f.) and average of the per-source AoI; iv) and analyse the average per-source AoI versus all system parameters.
The main contributions of this work are summarized as follows.
%\begin{itemize}
%  \item extend the 2D age process approach into the considered multi-source system, and present the detailed analyses on the per-source AoI evolution based on the Markovity of the 2D age process;
%  \item derive the analytical expressions of the probability mass function (p.m.f.) and average of the per-source AoI, and analyse the average per-source AoI versus all the system parameters.
%\end{itemize}
\begin{itemize}
  \item We extend the 2D age process approach into the discrete-time multi-source queueing system, and present the complete and detailed analyses on the per-source AoI evolution based on the Markovity of the 2D age process.
  \item The analytical expressions of the p.m.f. and average of the per-source AoI are derived. The AoI performance versus all system parameters and the effect of the retransmission protocol
%and preemption policy
are
analysed by numerical results.
\end{itemize}
%i) extend the 2D age process approach into the considered multi-source system, and present the detailed analyses on the per-source AoI evolution based on the Markovity of the 2D age process;
%ii) derive the analytical expressions of the probability mass function (p.m.f.) and average of the per-source AoI; iii) and analyse the average per-source AoI versus all the system parameters as well as the effects of the retransmission protocol and preemption queueing policy on the information freshness.
%, where the first element of the 2D age process is defined as the instantaneous per-source AoI while the second element represents whether an update of the concerned source is being transmitted and its current age;
%and present a complete framework to strictly analyse the per-source AoI process by using the
Specifically, similar to the approach in single-source systems, we define the first element of the 2D age process as the per-source AoI process.
To fit the 2D age process approach into the multi-source bufferless preemptive system and simplify the analysis,
% as much as possible,
we let the second element represent whether an update of the concerned source is in transmission and its current age, while ignoring the link details when no concerned-source update is being transmitted.
%either whether the transmitter is idle or which source's update is being transmitted when no update of the concerned source is being transmitted.
%这句话可以用来加上非抢占的问题之后再往上写。
%This overlook
%Specifically, to fully track the per-source AoI evolution, we define the first element of the 2D age process as the instantaneous per-source AoI, as well as utilize the second element to represent whether an update of the concerned source is being transmitted and its current age.
%Different from the definition of the 2D age process for the single-source system in \cite{9965796,9611393},
The Markovity of the 2D age process helps overcome the challenge for analysing the frequently disturbed queuing process caused by the preemption among sources.
%The 2D age process is a DTMC.
%These overcome the challenges caused by the preemption-leaded disturbed queuing process.
By analysing state transitions of the 2D age process, we further establish the stationary equations, through which the stationary distribution of the 2D age process is derived. This lays the foundation to find the p.m.f. and average of per-source AoI.
Numerical results validate the accuracy of the theoretical analyses and the effectiveness of the retransmission protocol, as well as analyse the average per-source AoI versus update generation and transmission success probabilities.
% and preemption policy.

The remaining part of this paper is organized as follows.
Section II introduces the system model.
In Section III, we analyse the per-source AoI process based on the 2D age process, and derive the analytical expressions of p.m.f. and average of the per-source AoI.
Numerical results and simulations are shown in Section IV. Finally, Section V concludes this paper.

\section{System Model}
%\vspace{-1 mm}
%\subsection{System Model}
%%就直接说一个源节点，源节点产生N类更新就好了
%\begin{figure}[!t]
%\centering\includegraphics[width=0.7\linewidth]{V2I.png}
%\caption{\textcolor{black}{The considered V2I link.}}
%%State transition diagram.
%%State transitions of $(n,m)$, $4\geq n>m\geq0$, $n\geq2$, as examples.}}
%\label{v2i}
%\end{figure}
Consider a V2I link where a vehicle transmits multiple kinds of vehicle status data to a RI.
%It is considered that t
The vehicle collects the status updates from $N$ information sources, e.g., velocity, acceleration, and surrounding object sensors.
%This system can be extracted as a multi-source status update system consisting of
%%or V2I uplink, i.e., wireless multi-source status updating system consisting of
%$N$ information sources (\emph{multiple sources}), a transmitter (\emph{single server}) and a monitor.
Each source independently generates respective updates at random instants, which would be sent to the RI by a transmitter via a wireless channel.
%It is assumed that the updates from different sources belong to different classes.
% and sends them to the monitor by utilizing the transmitter.
In particular, a discrete (slotted)-time model is adopted, where the time is divided into equal-length slots\cite{9377627,9965796,9611393,9333607}.
%(\emph{discrete-time queueing system}).

Specifically, assume that an update of source $i\in\mathcal{N}:=\left\{{1,2,\ldots,N}\right\}$, can be generated at the end of each slot with update generation probability $q_i$,
%(\emph{Bernoulli arrival}),
and the updates generated at the end of the $U$-th slot
%are operated with
possess time stamp $U\in\mathcal{Z}^+:=\left\{{1,2,\ldots}\right\}$.
%Specifically, assume that each source probabilistically generates an update with time stamp $U$ at the end of the $U$-th slot, and the update generation process
%% (arrival process)
%of source $i$ is a Bernoulli process with rate $q_i$, where $i\in\mathcal{N}:=\left\{{1,2,\ldots,N}\right\}$, $U\in\mathcal{Z}^+:=\left\{{1,2,\ldots}\right\}$.
%% and $1>q_i>0$.
%This represents that an update of source $i$ can be generated at the end of each slot with update generation probability $q_i$.
%This represents that an update can be generated at the end of each slot with probability $p$
%, where a class-$i$ update can be generated with probability $p_i:=pq_i$, $\sum\nolimits_{i=1}^N{p_i}=p$.
It is also assumed that each update transmission starts at the beginning of a slot and completes right before the end of that slot.
%lasts for one slot, and is completed right before the end of a slot.
In fact, transmission failures, e.g., decoding errors would probably occur due to the channel fading\cite{woo}.
The transmission success probabilities of the updates of different sources can be different,
% due to the possibly different packet lengths of sources.
 since the information rates of sources can be different.
Let us denote the transmission success probability of the source-$i$ updates at each slot by $\gamma_i$.
Retransmission protocol is adopted to combat the channel fading.
Specifically, if an update transmission fails (succeeds) at a slot and no new update is generated at the end, the RI instantaneously feedbacks an NACK (ACK) and the vehicle retransmits the update (is in idle).
% (\emph{geometrically distributed service time}).
No matter whether there is a transmission failure, if new updates are generated at the end of a slot, the vehicle randomly and uniformly selects one of the newly generated updates and immediately starts to transmit it at the beginning of the next slot by following the preemption queuing policy
%(\emph{zero queue length; preemption queuing discipline})
\cite{9333607}.
One can see the transmitter is bufferless except a storage for the update being transmitted.
It can be found that the vehicle always sends the latest generated update to the RI and the update waiting time is zero, which hold the potential for improving the information freshness.
%In effect, the total transmission time (service time) of a successfully recieved source-$i$ update follows a Geometric distribution with expectation $1/\gamma_i$.

In effect, the considered V2I link can be extracted as a multi-source Ber/Geo/1/1 preemptive queueing system:
%Specifically,
i) The arrival process of source $i$ is a Bernoulli process with rate $q_i$, since that the updates are generated at the end of each slot with probability $q_i$;
ii) The service time is source-dedicated and geometrically distributed with parameter $\gamma_i$, since the retransmission protocol is adopted and the transmission success
probability is $\gamma_i$;
iii) The queue length is zero and the system capacity is one, since the bufferless transmitter is employed.
The AoI is adopted as the information freshness metric.
%Specifically, the per-source AoI is defined as
%the time elapsed since the latest successfully received update of the concerned source was generated.
Recall the definition of AoI mentioned in Section \ref{intro}.
Denote the generation-time stamp of the latest successfully received update of source $i$ at slot $t$ by $U_i(t)$.
The AoI of source $i$ at slot $t$ can be given by
%\footnote{It is valuable to note that the stationary and ergodic per-steam AoI process is considered in this work.}
\begin{align}
\Delta_i(t):=t-U_i(t), \;t\in\mathcal{Z}^+.
\end{align}
%It is valuable to note that
%:
%i) The AoI processes of all streams are mutually independent, since the sensors separately update respective information to the monitor by orthogonal channels;
We consider that the per-source AoI process is stationary and ergodic, as commonly assumed in \cite{9312180,8820073,tmd,woo,9965796,9611393,9377627,9333607}.
In fact, subsequent theoretical analyses and simulations confirm this assumption.
% in this work.

%the sources probabilistically generate the updates with time stamp $U$ at the end of the $U$-th slot, and the transmitter starts to transmit one of the new updates at the beginning of the next slot.
%A evolution example of the per-source AoI process is presented in Fig. 1.
%%图里要画出来抢占、最小是2、以及标出成功传输和错误传输

To comprehensively characterize the information freshness, let us pay attention to the (stationary) distribution, e.g., the p.m.f. of the per-source AoI:
\begin{align}
\pi_i(n):=\Pr\left\{\Delta_i=n\right\}, \;n\in\mathcal{Z}^+ \backslash \{1\}.
\end{align}
Moreover, let us adopt average AoI to evaluate the information freshness effectively. The average AoI of source $i$ is given by
\begin{align}
\overline\Delta_i:=\mathbb{E}\left[{\Delta_i}\right]=\sum\nolimits_{n=2}^\infty{n\pi_i(n)},\label{average}
\end{align}
in which $\mathbb{E}\left[{\cdot}\right]$ represents the expectation operator.
It is noteworthy that the minimum of the per-source AoI is two slots, as in \cite{9377627}. This is because that
i) the successfully received update with time stamp $U$, of which minimum total transmission time is one slot, is generated at the end of the $U$-th slot;
ii) and starts to be transmitted at the beginning of the next slot.

\section{P.M.F. and Average of the Per-Source AoI}
%In this section, the distributions of per-stream AoI and per-stream PAoI in the system where each sensor is equipped with the classical ARQ protocol are fisrt obtained.
In this section, we focus on characterizing the statistics of the per-source AoI process.
%the per-source AoI process is analysed.
In particular, analytical expressions of the p.m.f. and average of the per-source AoI are derived.
%Without loss of generality, we focus on the AoI process of source $i$.
%In this section, let us concisely derive the explicit expressions of the distributions and averages of AoSI and PAoSI in a typical two-stream system with the classical ARQ protocol.
%The detailed derivations for this two-stream case can provide references to the cases with other given numbers of streams and updating protocols.
%Then, we showcase the explicit expressions of the distributions for the two-stream case, based on which the average AoSI and PAoSI are further obtained to quantify the timeliness of the stalest information.
%First, let us the analyse the evolution of the per-class AoI process $\Delta_i(t)$.
%%下面注释的这些实际上可以写在intro里面，写在contribution那里。
First, a Markov 2D age process is introduced to fully track the per-source AoI process.
Then, the state transitions of the 2D age process are analysed, based on which the stationary equations are established.
By solving the stationary equations, we find the stationary probabilities of the 2D age process.
Based on the stationary probabilities, the p.m.f. and average of the per-source AoI are finally derived.
%\vspace{-2.5 mm}
\subsection{Definition of the 2D Age Process }
%\vspace{-1 mm}
To fully track the evolution of the per-source AoI process $\Delta_i(t)$, let us utilize a 2D age process, which is defined as $\boldsymbol{\Delta}_i(t)=(\Delta_i(t),\Delta'_i(t))$.
Let $\left\{\Delta'_i(t)=0\right\}$ represent the event that no update is being transmitted or an update of an other source is being transmitted.
%On the other hand, l
Let event $\left\{\Delta'_i(t)\geq1\right\}$ stand for the opposite, i.e., a source-$i$ update is being transmitted at slot $t$.
%Let $\left\{\Delta'_i(t)\geq1\right\}$ represent the event that an source-$i$ update is being transmitted at slot $t$.
%On the other hand, let event $\left\{\Delta'_i(t)=0\right\}$ stand for the opposite, i.e., there is no update is being transmitted or an update of an other source is being transmitted.
Moreover, for the case where $\Delta'_i(t)\geq1$, let the value of $\Delta'_i(t)$ denote the time elapsed since the update from source $i$ being transmitted was generated, i.e., the current age of the source-$i$ update being transmitted.
Denote the state vector of the 2D age process by $(n,m)$, where $n\in\left\{\mathcal{Z}^+\backslash \{1\}\right\}$, $m\in\left\{\mathcal{Z}^+\cup\{0\}\right\}$.
Note that by definition, $n>m$ always holds.
%In fact, one can find that the age process $\boldsymbol{\Delta}_i(t)$ is a two-dimensional Markov chain.
%\vspace{-2.5 mm}
\subsection{State Transitions of the 2D Age Process}
%\vspace{-1 mm}
In this subsection, we study the state transitions of the 2D age process $\boldsymbol{\Delta}_i(t)$ and the state transition probabilities.

%To be clarity, let us present the following definitions first.
%Define the source-$i$ effective update generation probability $p_i$ as
% the probability that a source-$i$ update is generated and selected at the end of each slot, i.e.,
%\begin{align}
%p_i:=&\sum\nolimits_{h=0}^{N-1}\Bigg(\frac{q_i}{h+1}\sum\nolimits_{\mathcal{H}_h\subset\{\mathcal{N}\backslash\{i\}\}}\Bigg(\prod\limits_{j\in\mathcal{H}_h}q_j\nonumber\\
%&\times{\prod\limits_{l\in\{{\mathcal{N}\backslash
%\{{\mathcal{H}_h\bigcup\{i\}}\}}\}}(1-q_l)}\Bigg)\Bigg),
%\end{align}
%where $\mathcal{H}_h$ represents a $h$-ary subset of set $\mathcal{N}$, $h\in\mathcal{N}$.
%%Denote the probability that
%Define the overall update generation probability $p$ as the probability that one update or multiple update are generated at the end of each slot, i.e.,
%\begin{align}
%p:=1-\prod\limits_{i=1}^{N}(1-q_i).\label{444}
%\end{align}
%In fact, according to the definitions of $p_i$ and $p$, it has that $p=\sum\nolimits_{i=1}^Np_i$.
%Based on (\ref{444}), the effective update generation probability of source $i$ can be simplified as
%\begin{align}
%p_i=\frac{q_i(1-p)}{1-q_i}\sum\nolimits_{h=0}^{N-1}\Bigg(\frac{1}{h+1}\sum\nolimits_{\mathcal{H}_h\subset\{\mathcal{N}\backslash\{i\}\}}\Bigg({\prod\limits_{j\in\mathcal{H}_h}
%\frac{q_j}{1-q_j}}\Bigg)\Bigg).
%\end{align}

First, let us consider the case where no update of source $i$ is being transmitted at the current slot $\tau$.
One can assume that $\boldsymbol{\Delta}_i(\tau)=(n-1,0)$, where $n\geq3$.
In this case, the state transition of the 2D age process depends on whether an update of source $i$ is generated and selected by the vehicle at the end of the current slot.
%, since the preemption policy is adopted.
If an update of source $i$ is generated and selected at the end of slot $\tau$, this update will be transmitted at the beginning of the next slot, which leads to that $\boldsymbol{\Delta}_i(\tau+1)=(n,1)$. Otherwise, one can obtain that $\boldsymbol{\Delta}_i(\tau+1)=(n,0)$.
Define the effective update generation probability $p_i$ as the probability that an update of source $i$ is generated and selected at the end of each slot.
Recall that multiple updates of different sources can be generated at the same time, and the vehicle selects one of the newly generated updates randomly and uniformly to transmit.
The probability $p_i$ is given by
\begin{align}
p_i
:=\sum\nolimits_{h=0}^{N-1}\!\sum\limits_{\mathcal{H}_h\subset\{\mathcal{N}\backslash\{i\}\}}\!\frac{q_i\prod\nolimits_{j\in\mathcal{H}_h}q_j}{h+1}
{\prod\limits_{ l\in\{\mathcal{N}\backslash
\{\mathcal{H}_h \cup\{i\}\}\}}\!(1-q_l)}
=\frac{q_i(1-p)}{1-q_i}\sum\nolimits_{h=0}^{N-1}\sum\limits_{\mathcal{H}_h\subset\{\mathcal{N}\backslash\{i\}\}}{\frac{1}{h+1}\prod\limits_{j\in\mathcal{H}_h}
\frac{q_j}{1-q_j}},\label{pi}
\end{align}
where $\mathcal{H}_h$ stands for a $h$-ary subset of set $\mathcal{N}$ and $p$ represents the overall update generation probability, i.e.,
\begin{align}
p:=\sum\nolimits_{i=1}^Np_i=1-\prod\nolimits_{i=1}^{N}(1-q_i).\label{444}
\end{align}
%In fact, according to the definitions of $p_i$ and $p$, it has that $p=\sum\nolimits_{i=1}^Np_i$.
Accordingly, the state transition probabilities of the 2D age process in this case can be given by
\begin{align}
P_{(({n-1,0}),(n,1))}^{(i)}=&p_i,n\geq3,\label{a1}\\
P_{(({n-1,0}),(n,0))}^{(i)}=&1-p_i,n\geq3,\label{a2}
\end{align}
in which the first and second elements of the subscript of the state transition probability stands for the states of $\boldsymbol{\Delta}_i(t)$ at the current and next slots respectively.
%in which $n\geq3$.

Then, let us consider the case where an update of source $i$ is being transmitted at the current slot $\tau$.
%Five kinds of state transitions exist in this case.
Assume that $\boldsymbol{\Delta}_i(\tau)=(n-1,m-1)$, where $n> m\geq 2$.
The state transition in this case depends on whether the update being transmitted is successfully transmitted right before the end of the current slot, as well as whether a new update of source $i$ is generated and selected at the end of the slot.
%Assume that $\boldsymbol{\Delta}_i(\tau)=(n-1,m-1)$, where $n> m\geq 2$ and $n\geq 3$.
%\begin{itemize}
%  \item
%  \item
%  \item
%  \item
%  \item
%\end{itemize}
If the source-$i$ update is successfully transmitted and an update of source $i$ is generated and selected at the end of the slot, the new update starts to be transmitted at the beginning of the next slot. This leads to that $\boldsymbol{\Delta}_i(\tau+1)=(m,1)$.
If the update is successfully transmitted and no update of source $i$ is generated or selected, the 2D age process at the next slot is given by $\boldsymbol{\Delta}_i(\tau+1)=(m,0)$.
If the update transmission fails and an update of source $i$ is generated and selected at the end of the slot, the vehicle starts to transmit the new update instead at the beginning of the next slot. Hence, $\boldsymbol{\Delta}_i(\tau+1)=(n,1)$.
If the update transmission fails and an update of an other source is generated and selected, the transmission of the source-$i$ update is preempted and the new update is transmitted instead. This leads to that $\boldsymbol{\Delta}_i(\tau+1)=(n,0)$.
Finally, if the update transmission fails and no update is generated, the vehicle retransmits the update at the next slot. One can obtain that $\boldsymbol{\Delta}_i(\tau+1)=(n,m)$.
%Recall that the effective update generation probability and transmission success probability of source $i$, as well as the overall update generation probability are given by $p_i$, $\gamma_i$ and $p$ respectively.
Accordingly, the corresponding state transition probabilities are given by
\begin{align}
P_{(({n-1,m-1}),(m,1))}^{(i)}=&\gamma_ip_i,n> m\geq 2,\label{a3}\\
P_{(({n-1,m-1}),(m,0))}^{(i)}=&\gamma_i(1-p_i),n> m\geq 2,\label{a4}\\
P_{(({n-1,m-1}),(n,1))}^{(i)}=&(1-\gamma_i)p_i,n> m\geq 2,\label{a5}\\
P_{(({n-1,m-1}),(n,0))}^{(i)}=&(1-\gamma_i)(p-p_i),n> m\geq 2,\label{a6}\\
P_{(({n-1,m-1}),(n,m))}^{(i)}=&(1-\gamma_i)(1-p),n> m\geq 2.\label{a7}
\end{align}
%where $n> m\geq 2$ and $n\geq 3$.

\begin{figure}[!t]
\centering\includegraphics[width=0.4\linewidth]{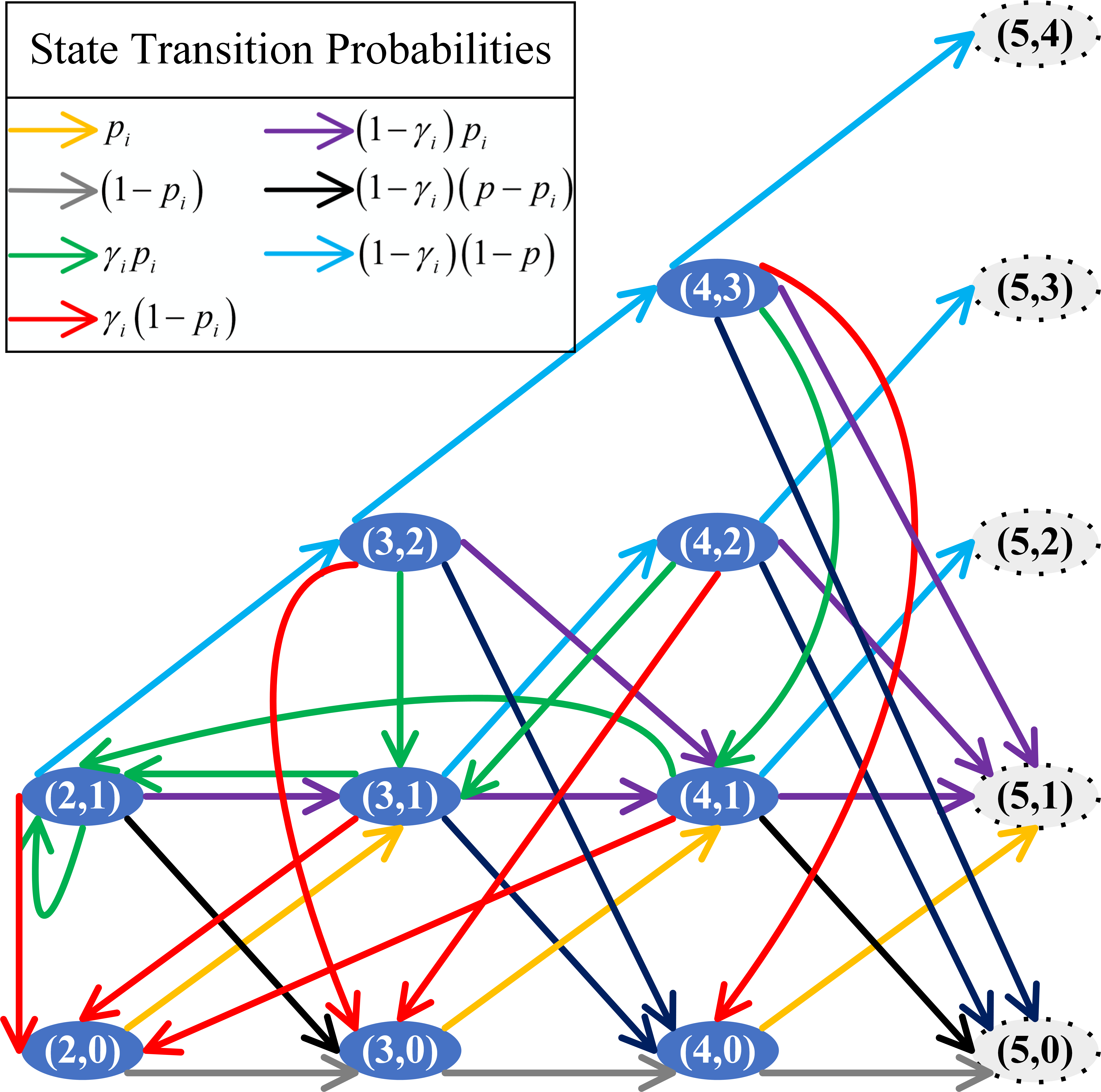}
\caption{\textcolor{black}{State transition diagram of the 2D age process. State transitions of states $(n,m)$ where $4\geq n>m\geq0$ and $n\geq2$, are shown as examples.}}
%State transition diagram.
%State transitions of $(n,m)$, $4\geq n>m\geq0$, $n\geq2$, as examples.}}
\label{zhuanyi}
\end{figure}
Based on the state transition analysis, it is found that the state of 2D age process at the next slot is only determined by the state at the current slot, regardless of the historical states.
%; ii) All the state transition probabilities are independent with time.
%One can deduce that
This indicates that
%the  age process
$\boldsymbol{\Delta}_i(t)$
is a
%time-homogeneous
2D DTMC.
To be clarity, a schematic view of the state transitions are presented in Fig. \ref{zhuanyi}.
%In fact, one can find that age process $\boldsymbol{\Delta}_i(t)$ is a two-dimensional Markov process.
%For $\Delta'_i(t)\geq1$, process $\Delta'_i(t)$
%\vspace{-1.75 mm}
\subsection{Stationary Equations of the 2D Age Process}
%\vspace{-1 mm}
Let us establish the stationary equations of 2D age process $\boldsymbol{\Delta}_i(t)$.
Denote the stationary probabilities of $\boldsymbol{\Delta}_i(t)$ by $\pi^{(i)}_{(n,m)}$.
%, where $n>m$, $n\in\left\{\mathcal{Z}^+\backslash \{1\}\right\}$ and $m\in\left\{\mathcal{Z}^+\cup\{0\}\right\}$.

Note that all the states of $\boldsymbol{\Delta}_i(t)$ can be divided into five classes, i.e., $(2,1)$, $(2,0)$, $(n,1)$, $(n,0)$ and $(n,m)$, where $n> m\geq2$.
According to the state transitions (cf. Fig. \ref{zhuanyi}) and the transition probabilities given by Eq.s (\ref{a1}), (\ref{a2}), (\ref{a3}), (\ref{a4}), (\ref{a5}), (\ref{a6}) and (\ref{a7}), it is found that:
i) Only state $(n,1)$, $n\geq2$, can transit to states $(2,1)$ and $(2,0)$ with probabilities $\gamma_i p_i$ and $\gamma_i(1-p_i)$ respectively;
ii) State $(n,1)$, $n\geq 3$, can be transited only from states $(n-1,0)$, $(k,n-1)$ for $k\geq n$, and $(n-1,k)$ for $n-1> k\geq1$, with probabilities $p_i$, $\gamma_i p_i$ and $(1-\gamma_i)p_i$ respectively;
iii) State $(n,0)$, $n\geq 3$, can be transited only from states $(n-1,0)$, $(k,n-1)$ for $k\geq n$, and $(n-1,k)$ for $n-1> k\geq1$, with probabilities $1-p_i$, $\gamma_i (1-p_i)$ and $(1-\gamma_i)(p-p_i)$ respectively;
iv) State $(n,m)$, $n> m\geq2$, can be transited only from state $(n-1,m-1)$ with probability $\lambda_i:=(1-\gamma_i)(1-p)$.
Hence, the stationary equations of $\boldsymbol{\Delta}_i(t)$ can be formulated as
\setcounter{equation}{13}
\renewcommand\theequation{\arabic{equation}}
\begin{align}
%&\text{Equations}\;\mathcal{E}_1:\label{P2}\\
\pi^{(i)}_{(2,1)}=&\gamma_i p_i \sum\nolimits_{n=2}^\infty\pi^{(i)}_{(n,1)},\tag{\theequation{a}}\label{P2a}\\
\pi^{(i)}_{(2,0)}=&\gamma_i (1-p_i) \sum\nolimits_{n=2}^\infty\pi^{(i)}_{(n,1)},\tag{\theequation{b}}\label{P2b}\\
\pi^{(i)}_{(n,1)}=&p_i\pi^{(i)}_{(n-1,0)}+\gamma_i p_i\sum\nolimits_{k=n}^\infty\pi^{(i)}_{(k,n-1)}+(1-\gamma_i) p_i\sum\nolimits_{k=1}^{n-2}\pi^{(i)}_{(n-1,k)},n\geq3,\tag{\theequation{c}}\label{P2c}\\
\pi^{(i)}_{(n,0)}=&(1-p_i)\pi^{(i)}_{(n-1,0)}+\gamma_i(1- p_i)\sum\nolimits_{k=n}^\infty\pi^{(i)}_{(k,n-1)}+(1-\gamma_i)(p-p_i)\sum\nolimits_{k=1}^{n-2}\pi^{(i)}_{(n-1,k)},n\geq3,\tag{\theequation{d}}\label{P2d}\\
\pi^{(i)}_{(n,m)}=&\lambda_i\pi^{(i)}_{(n-1,m-1)},n>m\geq2.\tag{\theequation{e}}\label{P2e}
\end{align}
%in which $\lambda_i:=(1-\gamma_i)(1-p)$.
%\vspace{-1.75 mm}
\subsection{Stationary Probabilities of the 2D Age Process as well as P.M.F. and Average of the Per-Source AoI}
%\vspace{-1 mm}
In this subsection, the stationary probabilities of the 2D age process $\boldsymbol{\Delta}_i(t)$ are found, based on which the p.m.f. and average of the per-source AoI ${\Delta}_i(t)$ are derived.

Based on the stationary equations of $\boldsymbol{\Delta}_i(t)$ formulated as
%$\mathcal{E}_1$,
Eq.s (\ref{P2a})--(\ref{P2e}),
the stationary probabilities can be obtained.
%, as presented in the following theorem.
\begin{lem}\label{thm1}
The stationary probabilities of the 2D age process $\boldsymbol{\Delta}_i(t)$ are given by
\begin{align}
\pi^{(i)}_{(n,0)}=&p_i\left({\frac{(\beta_i+\gamma_i-1)\beta^{n-1}_i-(\alpha_i+\gamma_i-1)\alpha^{n-1}_i}{\beta_i-\alpha_i}-\lambda^{n-1}_i}\right),n\geq2,\label{thq1}\\
\pi^{(i)}_{(n,m)}=&\gamma_i p^2_i\lambda^{m-1}_i\left({\frac{\beta^{n-m}_i-\alpha^{n-m}_i}{\beta_i-\alpha_i}}\right), n>m\geq1,\label{thq2}
\end{align}
where $p_i$ is given by Eq. (\ref{pi}), $\lambda_i=(1-\gamma_i)(1-p)$, as well as $\alpha_i$ and $\beta_i$ are the two different solutions of the quadratic equation $x^2-(1-\gamma_ip_i+\lambda_i)x+\lambda_i=0$ w.r.t. $x$.
\end{lem}
The proof of Lemma 1 is given in Appendix \ref{111}.

%\begin{proof}
%See Appendix
%\end{proof}
%\begin{figure*}[ht]
%  \centering
%
%  \subfigure[{The p.m.f. of per-source AoI in the system with retransmission protocol.}]{
%     \label{fenbu} %% label for first subfigure
%    \includegraphics[width=0.32\linewidth]{lisanduoliufenbu.eps}}\hspace*{\fill}
%  \subfigure[{The average  per-source AoI in the system with retransmission protocol.}]{
%    \label{junzhi} %% label for second subfigure
%    \includegraphics[width=0.32\linewidth]{lisanduoliujunzhi.eps}}\hspace*{\fill}
%    %\hspace*{\fill}
%  \subfigure[{The average per-source AoIs in the systems with and without retransmission protocol.}]{
%    \label{chongchuan} %% label for second subfigure
%    \includegraphics[width=0.32\linewidth]{youwuchongchuan.eps}}
%%  \subfigure[{The average overall AoI.}]{
%%    \label{zhengti} %% label for second subfigure
%%    \includegraphics[width=0.32\linewidth]{zhengtiaoijunzhi.eps}}
%%\vspace{-3 mm}
%  \caption{The p.m.f. and average of the per-source AoI in the two-source system.}
%%  \caption{The p.m.f. and average of the per-source AoI as well as the average overall AoI in the typical two-source system.}
%%  \vspace{-5 mm}
%  \label{fenbuhejunzhi} %% label for entire figure
%\end{figure*}
Note that the p.m.f. of the per-source AoI ${\Delta}_i(t)$ can be expressed by the stationary probabilities of $\boldsymbol{\Delta}_i(t)$, as
 \begin{align}
 \pi_i(n)=&\sum\nolimits_{m=0}^{n-1}\Pr\left\{\Delta_i=n,\Delta'_i=m\right\}
 =\pi^{(i)}_{(n,0)}+\sum\nolimits_{m=1}^{n-1}\pi^{(i)}_{(n,m)},n\geq2.\label{34}
 \end{align}
Based on Eq. (\ref{34}) and Lemma \ref{thm1},  the p.m.f. and average of per-source AoI can be derived, as presented in the following.
%According to the relationship between the p.m.f. of ${\Delta}_i(t)$ and the stationary probabilities of $\boldsymbol{\Delta}_i(t)$:
% \begin{align}
% \pi_i(n)=&\sum\nolimits_{m=0}^{n-1}\Pr\left\{\Delta_i=n,\Delta'_i=m\right\}\nonumber\\
% =&\pi^{(i)}_{(n,0)}+\sum\nolimits_{m=1}^{n-1}\pi^{(i)}_{(n,m)},n\geq2,\label{34}
% \end{align}
% the p.m.f. of per-source AoI and the average per-source AoI can be derived, as presented in the following.
\begin{thm}\label{thm2}
The p.m.f. and average of the per-source AoI ${\Delta}_i(t)$ are respectively given by
\begin{align}
 \pi_i(n)=&\gamma_ip_i\left({\frac{\beta^{n-1}_i-\alpha^{n-1}_i}{\beta_i-\alpha_i}}\right),n\geq2,\label{thqq1}\\
%\overline\Delta_i=&\frac{\gamma_i+p-\gamma_i(p-p_i)}{\gamma_ip_i},\label{thqq2}
\overline\Delta_i=&\frac{\gamma_i+(1-\gamma_i)p}{\gamma_ip_i}+1,\label{thqq2}
%\overline{\Delta}_i=&\frac{1}{\gamma_i}+\frac{1}{p_i}+\frac{(1-\gamma_i)p-p_i}{p_i}+1,\label{thqq2}
\end{align}
in which $p_i$ is given by Eq. (\ref{pi}), $p$ is given by Eq. (\ref{444}), as well as $\alpha_i$ and $\beta_i$ are the two different solutions of the quadratic equation $x^2-(1-\gamma_ip_i+\lambda_i)x+\lambda_i=0$ w.r.t. $x$.
%in which $p_i$ is given by (\ref{pi}), $\lambda_i=(1-\gamma_i)(1-p)$, as well as $\alpha_i$ and $\beta_i$ are the two different solutions of the quadratic equation $x^2-(1-\gamma_ip_i+\lambda_i)x+\lambda_i=0$ w.r.t. variable $x$.
\begin{proof}
First, the p.m.f. of per-source AoI given by Eq. (\ref{thqq1}) can be obtained by substituting Eq.s (\ref{thq1}) and (\ref{thq2}) into Eq. (\ref{34}) as well as using Eq. (\ref{26}).
On the other hand, one can obtain the average per-source AoI given by Eq. (\ref{thqq2}) by plugging Eq. (\ref{thqq1}) into
Eq. (\ref{average})
as well as utilizing $0<\alpha_i<1$, $0<\beta_i<1$, Eq. (\ref{26}) and $\lambda_i=(1-\gamma_i)(1-p)$.
It is noteworthy that Eq. (\ref{26}) as well as the proofs of $0<\alpha_i<1$ and $0<\beta_i<1$ are presented in Appendix \ref{111}.
The detailed computations in this proof are omitted due to the page limit.
\end{proof}
\end{thm}
By setting $q_j=0$, $j\in\mathcal{N}$, $j\neq i$, one can find that Theorem \ref{thm2} is able to be applied to the single-source case, where the results in this work agree with those in \cite{9377627}.
%\begin{rem}
%Note that based on Eq. (\ref{thqq2}) and that $p=\sum\nolimits_{j=1}^Np_j$, the average per-source AoI can be rewritten as
%\begin{align}
%\overline{\Delta}_i=\frac{1}{\gamma_i}+\frac{1}{p_i}+\left({\frac{1-\gamma_i}{\gamma_i}}\right)\frac{\sum\nolimits_{j\in\{\mathcal{N}\backslash\{i\}\}}p_j}{p_i}.\label{zuihou}
%%\frac{1-p}{p_i}+\frac{p}{\gamma_ip_i}+1=\frac{1-p}{p_i}+\frac{\sum\nolimits_{j\subset\{\mathcal{N}\backslash\{i\}\}}p_j}{\gamma_ip_i}+2.\nonumber
%\end{align}
%First, one can find that the average per-source AoI decreases with the transmission success probability of the concerned source and is regardless of those of the other sources.
%Moreover, Eq. (\ref{pi}) shows that $p_i$ increases with $q_i$ and decreases with $q_j$, $j\neq i$.
%By combining this and Eq. (\ref{zuihou}), it is found that the average per-source AoI decreases with the update generate probability of the concerned source and increases with those of other sources.
%At last, by setting $q_j=0$, $j\neq i$, it can be found that Theorem \ref{thm2} can be applied to the single-source case, where the results in this work agree with those in \cite{9377627}.
%%, since $p/p_i=1+(\sum\nolimits_{j\subset\{\mathcal{N}\backslash\{i\}\}}p_j)/p_i$.
%In summary, the per-source information freshness can be improved by increasing the source-dedicated transmission success probability and update generation probability.
%\end{rem}
%It is valuable to note that the explicit expressions of the p.m.f. and average of per-source AoI derived in this work can be extended to single-source case.
%\vspace{-2.5 mm}
\section{Numerical Results}
In this section, we first present and study the p.m.f. and average of per-source AoI.
Simulations are provided to validate the theoretical analyses.
Then, the effect of the retransmission protocol is analysed.
%Then, the effects of the retransmission protocol and the preemption queueing policy are analysed.
%The average per-source AoIs in the systems with and without the retransmission protocol are compared.
%Then, the average overall AoI is also analysed.
%For more insights, the typical two-source system is considered.

\begin{figure}[ht]
  \centering
  \hspace*{\fill}
  \subfigure[{The p.m.f. of per-source AoI.}]{
     \label{fenbu} %% label for first subfigure
    \includegraphics[width=0.4\linewidth]{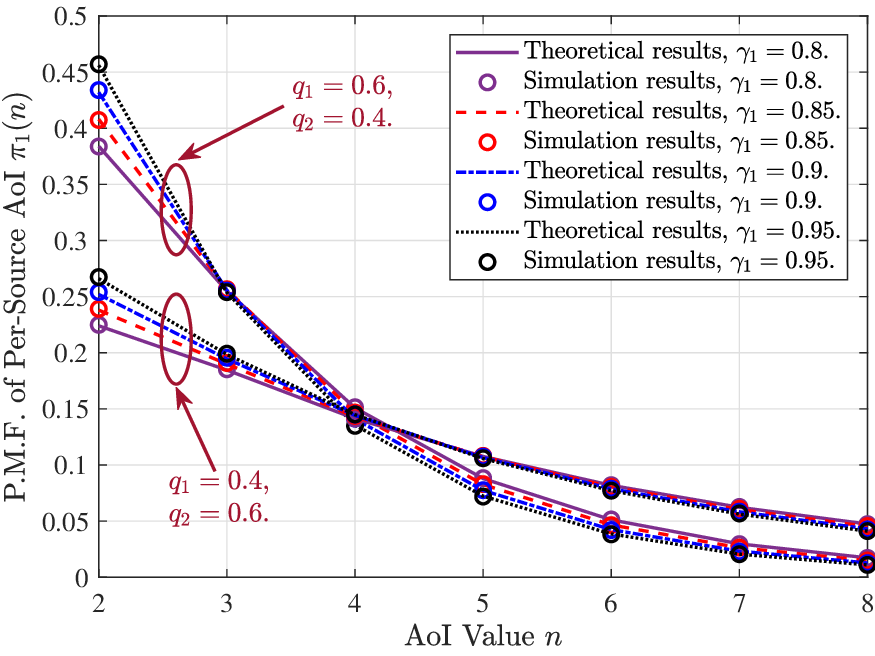}}
  \hspace*{\fill}
  \subfigure[{The average per-source AoI.}]{
    \label{junzhi} %% label for second subfigure
    \includegraphics[width=0.4\linewidth]{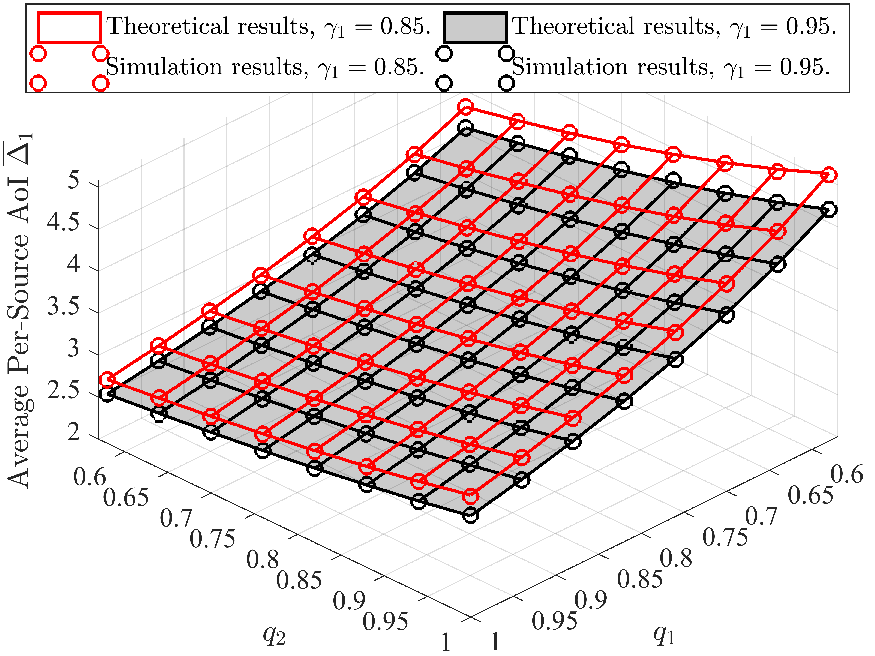}}
    \hspace*{\fill}
%  \subfigure[{The average overall AoI.}]{
%    \label{zhengti} %% label for second subfigure
%    \includegraphics[width=0.32\linewidth]{zhengtiaoijunzhi.eps}}
%\vspace{-3 mm}
  \caption{The p.m.f. and average of per-source AoI in the two-source V2I link.}
%  \caption{The p.m.f. and average of the per-source AoI as well as the average overall AoI in the typical two-source system.}
%  \vspace{-5 mm}
  \label{fenbuhejunzhi} %% label for entire figure
\end{figure}
%\subsection{P.M.F. and Average of the Per-Source AoI}
%Fig. \ref{fenbuhejunzhi} presents the p.m.f. and average of per-source AoI, where the AoI of source $1$ is focused on.
Fig. \ref{fenbu} presents the p.m.f. of per-source AoI versus transmission success probabilities.
%We simulate the per-source AoI processes in $1\times 10^5$ slots.
It is found that the probability distribution of per-source AoI tends to concentrate on the lower AoI value with the transmission success probability.
This is consistent with the intuition that the better communication conditions leads to the better  freshness.
Fig. \ref{junzhi} describes the average per-source AoI versus  update generation probabilities.
%The per-source AoI processes are simulated in $2\times 10^4$ slots.
It is shown that the average concerned-source AoI decreases with the update generation probability of the concerned source, however increases with that of the other source.
Because:
i) The increase of the concerned-source update generation probability raises the probability that the concerned-source update preempts the stale updates of all sources, which improves the concerned-source freshness;
ii) Oppositely, the increases of the update generation probabilities of the other sources raise the probability that the concerned source is preempted by the other sources, which leads to the deterioration of the information freshness of concerned source.
Besides, we simulate the per-source AoI processes in $10^5$ slots in Fig. \ref{fenbuhejunzhi}.
The simulation results verify the theoretical analyses.
%The coincidence between the simulation results and theoretical results validates the theoretical analyses.

\begin{figure}[!t]
\centering\includegraphics[width=0.4\linewidth]{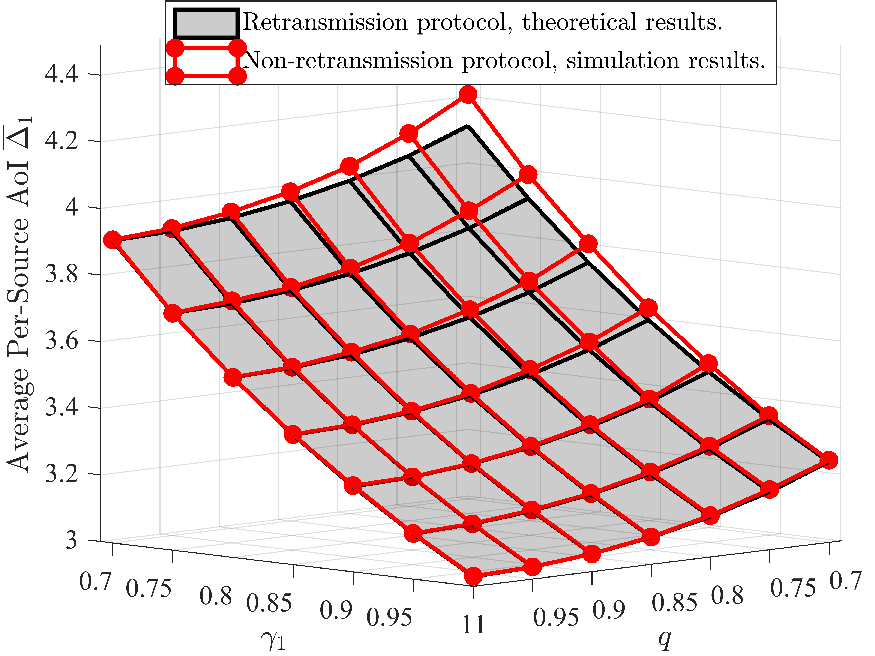}
\caption{\textcolor{black}{The average per-source AoIs in the two-source V2I links with and without the retransmission protocol.}}
%State transition diagram.
%State transitions of $(n,m)$, $4\geq n>m\geq0$, $n\geq2$, as examples.}}
\label{chongchuan}
\end{figure}
Fig. \ref{chongchuan} shows the average per-source AoIs in the V2I links with and without retransmission protocol.
Consider two V2I links where one of them is adopted with the retransmission protocol (i.e., the V2I link considered before) while the other is not, and that the update generation probabilities in the V2I links are equal to $q$.
In the latter V2I link, if an update transmission fails, this update is
%immediately
discarded.
%We simulate the per-source AoI process in this systems in $10^5$ slots.
It is presented that the AoI in the V2I link with retransmission is lower especially when the update generation and transmission success probabilities are low, which represents that the retransmission can improve the information freshness.
Because the retransmission enables the RI the chance to receive older but still fresh updates when no new update is generated at the vehicle.
Moreover, it is found that the retransmission-leaded freshness improvement, i.e., the effect of the retransmission protocol increases along with the decreases of the update generation and transmission success probabilities.
Because lower update generation and transmission success probabilities lead to higher probability of that the vehicle retransmits updates in the V2I link with retransmission while it is idle in the other V2I link.
\section{Conclusions}
Information freshness of
%the V2V or
the V2I link, which is extracted as the multi-source Ber/Geo/1/1 preemptive queueing system with the heterogeneous service time, was studied in this work.
%The information freshness of a wireless multi-source status updating system was studied in this work. The considered system was extracted as the multi-source Ber/Geo/1/1 preemptive queueing system.
To comprehensively characterize and evaluate the per-source AoI, we focused on the distribution and average.
Specifically, to fully track the per-source AoI process, we introduced a Markov 2D age process.
%, which was modeled as a Markov chain.
%The first element of the 2D age process is defined as the instantaneous per-source AoI, while the second element represents whether an update of concerned source is being transmitted and its current age.
By analysing the state transitions, state transitions probabilities and establishing the stationary equations, we found the stationary probabilities of the 2D age process.
Based on the stationary probabilities of the 2D age process, we derived the analytical expressions of the p.m.f. and average of the per-source AoI.
Finally, the numerical results validated the correctness of the theoretical analyses.
%\vspace{-0.6 mm}
\appendix
\subsection{Proof of Lemma \ref{thm1}}\label{111}
First, let us simplify the stationary equations of the 2D age process $\boldsymbol{\Delta}_i(t)$ given by Eq.s (\ref{P2a})--(\ref{P2e}).
Note that based on Eq.s (\ref{P2a})--(\ref{P2e}), one can find the following results:
\begin{align}
\pi^{(i)}_{(2,1)}=&\frac{p_i}{1-p_i}\pi^{(i)}_{(2,0)},\label{1}\\
\pi^{(i)}_{(n,m)}=&\lambda^{m-1}_i\pi^{(i)}_{(n-m+1,1)},n>m\geq1,\label{2}\\
\sum\nolimits_{k=n}^{\infty}\pi^{(i)}_{(k,n-1)}=&\sum\nolimits_{k=n}^{\infty}\lambda^{n-2}_i\pi^{(i)}_{(k-n+2,1)}
=\lambda^{n-2}_i\sum\nolimits_{h=2}^{\infty}\pi^{(i)}_{(h,1)}=\frac{\lambda^{n-2}_i}{\gamma_i(1-p_i)}\pi^{(i)}_{(2,0)},\label{3}\\
\sum\nolimits_{k=1}^{n-2}\pi^{(i)}_{(n-1,k)}=&\sum\nolimits_{k=1}^{n-2}\lambda^{k-1}_i\pi^{(i)}_{(n-k,1)}
=\lambda^{n-2}_i\sum\nolimits_{r=2}^{n-1}\lambda^{-r+1}_i\pi^{(i)}_{(r,1)},n\geq3.\label{4}
\end{align}
Specifically, i) Eq. (\ref{1}) holds following form Eq.s (\ref{P2a}) and (\ref{P2b}); ii) One can obtain Eq. (\ref{2}) by repeatedly iterating Eq. (\ref{P2e}); iii) Eq. (\ref{3}) holds following from Eq.s (\ref{P2a}) and (\ref{2}); iv) Eq. (\ref{4}) holds following from Eq. (\ref{2}). By substituting Eq.s (\ref{1}), (\ref{2}), (\ref{3}) and (\ref{4}) into $\mathcal{E}_1$, the stationary equations of $\boldsymbol{\Delta}_i(t)$ can be simplified as
\setcounter{equation}{23}
\renewcommand\theequation{\arabic{equation}}
\begin{align}
%&\text{Equations}\;\mathcal{E}_2:\label{P3}\\
\pi^{(i)}_{(2,1)}=&\gamma_i p_i \sum\nolimits_{n=2}^\infty\pi^{(i)}_{(n,1)},\tag{\theequation{a}}\label{P3a}\\
\pi^{(i)}_{(2,1)}=&\frac{p_i}{1-p_i}\pi^{(i)}_{(2,0)},\tag{\theequation{b}}\label{P3b}\\
\pi^{(i)}_{(n,1)}=&p_i\pi^{(i)}_{(n-1,0)}+\frac{p_i}{1-p_i}\lambda^{n-2}_i\pi^{(i)}_{(2,0)}
+(1-\gamma_i) p_i\lambda^{n-2}_i\sum\nolimits_{r=2}^{n-1}\lambda^{-r+1}_i\pi^{(i)}_{(r,1)},n\geq3,\tag{\theequation{c}}\label{P3c}\\
\pi^{(i)}_{(n,0)}=&(1-p_i)\pi^{(i)}_{(n-1,0)}+\lambda^{n-2}_i\pi^{(i)}_{(2,0)}+(1-\gamma_i)
(p-p_i)\lambda^{n-2}_i\sum\nolimits_{r=2}^{n-1}\lambda^{-r+1}_i\pi^{(i)}_{(r,1)},n\geq3,\tag{\theequation{d}}\label{P3d}\\
\pi^{(i)}_{(n,m)}=&\lambda^{m-1}_i\pi^{(i)}_{(n-m+1,1)},n>m\geq1.\tag{\theequation{e}}\label{P3e}
\end{align}

%To obtain the stationary probabilities of $\boldsymbol{\Delta}_i(t)$, l
Let us solve Eq.s (\ref{P3a})--(\ref{P3e}).
By observing, it is found that $\pi^{(i)}_{(n,0)}$ can be derived first.
It is valuable to note that
\begin{align}
\frac{1}{p_i}\left({\pi^{(i)}_{(n,1)}-p_i\pi^{(i)}_{(n-1,0)}-\frac{p_i}{1-p_i}\lambda^{n-2}_i\pi^{(i)}_{(2,0)}}\right)
=\frac{1}{p-p_i}\left({\pi^{(i)}_{(n,0)}-(1-p_i)\pi^{(i)}_{(n-1,0)}-\lambda^{n-2}_i\pi^{(i)}_{(2,0)}}\right),n\geq 3,
\end{align}
which holds following from Eq.s (\ref{P3c}) and (\ref{P3d}). One can deduce the relationship between $\pi^{(i)}_{(n,0)}$ and $\pi^{(i)}_{(n,1)}$, $n\geq3$, as
\begin{align}
\pi^{(i)}_{(n,1)}=&\frac{p_i}{p-p_i}\left({\pi^{(i)}_{(n,0)}-(1-p)\pi^{(i)}_{(n-1,0)}-\frac{1-p}{1-p_i}\lambda^{n-2}_i\pi^{(i)}_{(2,0)}}\right),n\geq3.\label{20}
\end{align}
Based on Eq.s (\ref{P3d}) and (\ref{20}), one can respectively obtain that
\begin{align}
(1-\gamma_i)(p-p_i)\lambda^{n-3}_i\sum\nolimits_{r=2}^{n-2}\lambda^{-r+1}_i\pi^{(i)}_{(r,1)}
=&\pi^{(i)}_{(n-1,0)}-(1-p_i)\pi^{(i)}_{(n-2,0)}-\lambda^{n-3}_i\pi^{(i)}_{(2,0)},n\geq4,\label{21}\\
\pi^{(i)}_{(n-1,1)}=&\frac{p_i}{p-p_i}\left({\pi^{(i)}_{(n-1,0)}-(1-p)\pi^{(i)}_{(n-2,0)}-\frac{1-p}{1-p_i}\lambda^{n-3}_i\pi^{(i)}_{(2,0)}}\right),n\geq4.\label{22}
\end{align}
On the other hand, Eq. (\ref{P3d}) can be rewritten as
\begin{align}
\pi^{(i)}_{(n,0)}
\!=\!(1-p_i)\pi^{(i)}_{(n-1,0)}\!+\!\lambda^{n-2}_i\pi^{(i)}_{(2,0)}\!+\!(1-\gamma_i)(p-p_i)\pi^{(i)}_{(n-1,1)}
\!+\!\lambda_i(1-\gamma_i)(p-p_i)\lambda^{n-3}_i\sum\nolimits_{r=2}^{n-2}\lambda^{-r+1}_i\pi^{(i)}_{(r,1)},n\geq3.\label{23}
\end{align}
By substituting Eq.s (\ref{21}) and (\ref{22}) into Eq. (\ref{23}) as well as utilizing $\lambda_i=(1-\gamma_i)(1-p)$, the second-order recursive expression of $\pi^{(i)}_{(n,0)}$ can be deduced, as
\begin{align}
\pi^{(i)}_{(n,0)}=(1-\gamma_ip_i+\lambda_i)\pi^{(i)}_{(n-1,0)}-\lambda_i\pi^{(i)}_{(n-2,0)}-\frac{p_i}{1-p_i}\lambda^{n-2}_i\pi^{(i)}_{(2,0)}, n\geq4.\label{24}
\end{align}
Let us derive $\pi^{(i)}_{(n,0)}$ by decreasing the order of the recursion. It is valuable to note that Eq. (\ref{24}) can be rewritten as
\begin{align}
\pi^{(i)}_{(n,0)}-\alpha_i\pi^{(i)}_{(n-1,0)}=&\beta_i\left({\pi^{(i)}_{(n-1,0)}-\alpha_i\pi^{(i)}_{(n-2,0)}}\right)-\frac{p_i}{1-p_i}\lambda^{n-2}_i\pi^{(i)}_{(2,0)}, n\geq4,\label{25}
\end{align}
in which $\alpha_i$ and $\beta_i$ satisfy that
\begin{align}
\begin{cases}
\alpha_i+\beta_i=1-\gamma_ip_i+\lambda_i,\\
\alpha_i\beta_i=\lambda_i,
\end{cases}\label{26}
\end{align}
i.e., $\alpha_i$ and $\beta_i$ are the two different solutions of the quadratic equation $x^2-(1-\gamma_ip_i+\lambda_i)x+\lambda_i=0$ w.r.t. $x$.
In fact, this quadratic equation always has two different solutions, since that the corresponding discriminant satisfies that
\begin{align}
(1-\gamma_ip_i+\lambda_i)^2-4\lambda_i
=((1-\gamma_i)+(1-p)+\gamma_i(p-p_i))^2-4(1-\gamma_i)(1-p)
>&(-(1-\gamma_i)+(1-p)+\gamma_i(p-p_i))^2\nonumber\\
\geq&0,
\end{align}
%which holds following form that $\lambda_i=(1-\gamma_i)(1-p)$.
where the first equality holds following from that $\lambda_i=(1-\gamma_i)(1-p)$.
By repeatedly iterating Eq. (\ref{25}), one can obtain the first-order recursive expression of $\pi^{(i)}_{(n,0)}$, as
\begin{align}
\pi^{(i)}_{(n,0)}-\alpha_i\pi^{(i)}_{(n-1,0)}=\beta^{n-3}_i\left({\pi^{(i)}_{(3,0)}-\alpha_i\pi^{(i)}_{(2,0)}}\right)-\frac{p_i}{1-p_i}\sum\nolimits_{k=0}^{n-4}{\beta^k_i\lambda^{n-k-2}_i}\pi^{(i)}_{(2,0)}, n\geq4.
\end{align}
Note that based on Eq. (\ref{P3d}) and $\lambda_i=(1-\gamma_i)(1-p)$, $\pi^{(i)}_{(3,0)}$ can be expressed as
\begin{align}
\pi^{(i)}_{(3,0)}=\left({\alpha_i+\beta_i-\frac{p_i\lambda_i}{1-p_i}}\right)\pi^{(i)}_{(2,0)}.
\end{align}
The first-order recursive expression can be simplified as
\begin{align}
\pi^{(i)}_{(n,0)}=\alpha_i\pi^{(i)}_{(n-1,0)}+\left({\left({1+\frac{p_i\lambda_i}{(1-p_i)(\lambda_i-\beta_i)}}\right)\beta^{n-2}_i-\frac{p_i\lambda_i}{(1-p_i)(\lambda_i-\beta_i)}\lambda^{n-2}_i}\right)\pi^{(i)}_{(2,0)},n\geq3.\label{28}
\end{align}
By repeatedly iterating Eq. (\ref{28}), $\pi^{(i)}_{(n,0)}$ can be determined by
\begin{align}
\pi^{(i)}_{(n,0)}
=&\alpha^{n-2}_i\pi^{(i)}_{(n-1,0)}+\left({\left({1+\frac{p_i\lambda_i}{(1-p_i)(\lambda_i-\beta_i)}}\right)\sum\nolimits_{k=0}^{n-3}\alpha^{k}_i\beta^{n-k-2}_i-\frac{p_i\lambda_i}{(1-p_i)(\lambda_i-\beta_i)}\sum\nolimits_{k=0}^{n-3}\alpha^{k}_i\lambda^{n-k-2}_i}\right)\pi^{(i)}_{(2,0)}\nonumber\\
=&\frac{1}{\gamma_i(1-p_i)}\left({\frac{(\beta_i+\gamma_i-1)\beta^{n-1}_i-(\alpha_i+\gamma_i-1)\alpha^{n-1}_i}{\beta_i-\alpha_i}-\lambda^{n-1}_i}\right){\pi^{(i)}_{(2,0)}},n\geq2,\label{29}
\end{align}
where the last equality holds following from Eq. (\ref{26}).
%Based on (\ref{29}) as well as the relationship between $\pi^{(i)}_{(n,1)}$ and $\pi^{(i)}_{(n,0)}$, one can determine $\pi^{(i)}_{(n,1)}$.
By substituting Eq. (\ref{29}) into
%the relationship between $\pi^{(i)}_{(n,1)}$ and $\pi^{(i)}_{(n,0)}$ given by
Eq. (\ref{20}), as well as utilizing $\lambda_i=(1-\gamma_i)(1-p)$ and Eq. (\ref{26}), one can further obtain $\pi^{(i)}_{(n,1)}$, as
\begin{align}
\pi^{(i)}_{(n,1)}=\frac{p_i}{1-p_i}\left({\frac{\beta^{n-1}_i-\alpha^{n-1}_i}{\beta_i-\alpha_i}}\right)\pi^{(i)}_{(2,0)},n\geq2.\label{30}
\end{align}
Furthermore, by plugging Eq. (\ref{30}) into Eq. (\ref{P3e}),
%i.e., the relationship between $\pi^{(i)}_{(n,m)}$ and $\pi^{(i)}_{(n,1)}$,
$\pi^{(i)}_{(n,m)}$, $n>m\geq1$, can be determined by
\begin{align}
\pi^{(i)}_{(n,m)}=\frac{p_i}{1-p_i}\lambda^{m-1}_i\left({\frac{\beta^{n-m}_i-\alpha^{n-m}_i}{\beta_i-\alpha_i}}\right)\pi^{(i)}_{(2,0)}.\label{31}
\end{align}
%in which $n>m\geq1$.

Based on Eq.s (\ref{29}) and (\ref{31}), it is found that to obtain the explicit expressions of the stationary probabilities, $\pi^{(i)}_{(2,0)}$ needs to be derived first. In effect, according to the regularization condition of the stationary probabilities, one can obtain that
\begin{align}
1=\sum\limits_{n>m\geq0,n\geq2}\pi^{(i)}_{(n,m)}
%=&\sum\nolimits_{n=2}^{\infty}\pi^{(i)}_{(n,0)}+\sum\nolimits_{n=2}^{\infty}\sum\nolimits_{m=1}^{n-1}\pi^{(i)}_{(n,m)}\nonumber\\
=\sum\nolimits_{n=2}^{\infty}\left({\pi^{(i)}_{(n,0)}+\sum\nolimits_{m=1}^{n-1}\pi^{(i)}_{(n,m)}}\right)
=\frac{\pi^{(i)}_{(2,0)}}{1-p_i}\sum\nolimits_{n=2}^{\infty}\frac{\beta^{n-1}_i-\alpha^{n-1}_i}{\beta_i-\alpha_i}
%=&\frac{\pi^{(i)}_{(2,0)}}{\gamma_i(1-p_i)}\sum\nolimits_{n=2}^{\infty}\left({\frac{(\beta_i+\gamma_i-1)\beta^{n-1}_i-(\alpha_i+\gamma_i-1)\alpha^{n-1}_i}{\beta_i-\alpha_i}}\right.\nonumber\\
%&\left.{-\lambda^{n-1}_i+\lambda^{n-1}_i-{\frac{(\beta_i-1)\beta^{n-1}_i-(\alpha_i-1)\alpha^{n-1}_i}{\beta_i-\alpha_i}}}\right)\nonumber\\
=\frac{\pi^{(i)}_{(2,0)}}{\gamma_ip_i(1-p_i)},\label{32}
\end{align}
where the third equality holds following from Eq.s (\ref{26}), (\ref{29}) and (\ref{31}), and the last equality holds based on that $0<\alpha_i<1$ and $0<\beta_i<1$ as well as Eq. (\ref{26}).
In particular, $0<\alpha_i<1$ and $0<\beta_i<1$ always hold since that
\begin{align}
\alpha_i\beta_i=&\lambda_i=(1-\gamma_i)(1-p)>0,\\
0<\alpha_i+\beta_i=&1-\gamma_ip_i+\lambda_i
%=&(1-\gamma_i)+(1-p)+\gamma_i(p-p_i)\nonumber\\
<(1-\gamma_i)(1-p)<1.
\end{align}
% $\alpha_i\beta_i=\lambda_i=(1-\gamma_i)(1-p)>0$ and $0<\alpha_i+\beta_i=1-\gamma_ip_i+\lambda_i=(1-\gamma_i)+(1-p)+\gamma_i(p-p_i)<(1-\gamma_i)(1-p)<1$.
Eq. (\ref{32}) implies that
\begin{align}
{\pi^{(i)}_{(2,0)}}={\gamma_ip_i(1-p_i)}.\label{33}
\end{align}

Finally, by plugging Eq. (\ref{33}) into Eq.s (\ref{29}) and (\ref{31}), all the stationary probabilities of $\boldsymbol{\Delta}_i(t)$ can be given by Eq.s (\ref{thq1}) and (\ref{thq2}). The detailed computations in this proof are omitted.
%The approach in this proof is elaborated, while the detailed computations are omitted.
%The detailed computations in this proof are omitted. Only the approach is presented.
%\begin{align}
%1=\sum\nolimits_{n>m\geq0,n\geq2}\pi^{(i)}_{(n,m)}=\sum\nolimits_{n=2}^{\infty}\pi^{(i)}_{(n,0)}+\sum\nolimits_{n=2}^{\infty}\sum\nolimits_{m=1}^{n-1}\pi^{(i)}_{(n,m)},
%\end{align}
%as well as utilizing $\lambda_i=(1-\gamma_i)(1-p)$ and (\ref{26}), $\pi^{(i)}_{(2,0)}$ can be given by
%\begin{align}
%\pi^{(i)}_{(2,0)}=\gamma_ip_i(1-p_i).
%\end{align}
\bibliographystyle{IEEEtran}
\bibliography{vpp}

\ifCLASSOPTIONcaptionsoff
  \newpage
\fi
\end{document}